\documentclass[twocolumn,showpacs]{revtex4}
\usepackage{graphicx}

\begin{document}
\title{Ultracold atoms confined in rf-induced two-dimensional trapping potentials}
\author{Yves Colombe, Elena Knyazchyan, Olivier Morizot, Brigitte Mercier, Vincent Lorent and H{\'e}l{\`e}ne Perrin}
\affiliation{Laboratoire de physique des lasers, CNRS-Universit{\'e} Paris 13\\
99 avenue Jean-Baptiste Cl{\'e}ment, F-93430 Villetaneuse}
\begin{abstract}
We present the experimental implementation of a new trap for cold atoms proposed by O.~Zobay and B.~M.~Garraway~\cite{Zobay01}. It relies on adiabatic potentials for atoms dressed by a rf field in an inhomogeneous magnetic field. This trap is well suited to confine atoms tightly along one direction to produce a two-dimensional atomic gas. We transferred ultracold atoms into this trap, starting either from thermal samples or Bose--Einstein condensates. In the latter case, technical noise during the loading stage caused heating and prevented us from observing 2D BECs.
\end{abstract}
\pacs{42.50.Vk, 03.75.-b, 32.80.Pj}
%
%
\maketitle
It is well known that Bose--Einstein condensation (BEC) of homogeneous gases cannot occur in dimensions less than 3 \cite{Hohenberg67}. However 2D condensation is possible inside a power law external confining potential \cite{Bagnato91}. Crossover from 3D condensation to 2D and 1D has been experimentally demonstrated by measuring the aspect ratios of atomic clouds released from anisotropic traps \cite{Ketterle01}. Quantum gases in low dimensions and particularly in 2D are of much interest for several reasons. It is predicted that the coherence and the mean-field interaction are strongly influenced by the trap parameters in the tightening direction \cite{Petrov00}. These properties are of crucial importance for integrated atom optics \cite{Zimmermann01,Reichel01}.
In 2D, the appearance of $1/2$-anyons quasiparticles has been predicted in a rotating trap \cite{Zoller01}. The first step towards the creation of these quasiparticles is to generate BEC in the lowest Landau level. This has been recently realized in a 3D rotating trap \cite{Cornell04}. By the centrifugal force the BEC approaches the two-dimensional regime.

There should be a dramatic improvement of this kind of experiments if the trapped gas is two-dimensional at the start. A 2D gas is obtained by strongly confining atoms along one direction. There are several ways to produce such a ``2D trap''. One is to use rapidly decaying fields supported by a surface, which provides very naturally a strong transverse confinement. The decaying field may be either optical, as for the double evanescent wave trap \cite{Ovchinnikov91,Hammes03}, or magnetic like in the Zeeman effect surface trap \cite{Hinds98}. 2D BEC was recently achieved in a trap combining the use of a single evanescent wave and gravity \cite{Grimm03}. However, the use of surface traps may lead to perturbation of the atomic cloud due for instance to the roughness of the trapping potential, as pointed out in \cite{Grimm03}. This effect is similar to the condensate fragmentation observed in atom chips \cite{fragmentation}. This is a disadvantage for experiments similar to \cite{Cornell04}. The use of an optical standing wave is a natural way to produce 2D traps far from any surface \cite{Bouchoule,Inguscio02}. The drawback of this method is that the atoms are filled in many pancake traps, unless the lattice period is chosen large enough. We present here the experimental implementation of a single pancake trap sitting in free space.

A new quasi 2D trap was proposed by Zobay and Garraway \cite{Zobay01,Zobay03}. It relies on the adiabatic potentials seen by an atom sitting in an inhomogeneous magnetic field $\mathbf{B_0}(\mathbf{r})$ and dressed by a radio-frequency field. The same potentials are implied in rf-induced evaporative cooling; in that case, they are used to limit the depth of the trap.

To realize the trapping potential, we start from the magnetic trap we use for producing a Bose--Einstein condensate. $^{87}$Rb atoms in the $5S_{1/2},F=2$ state sit in this inhomogeneous magnetic field and experience a $m_F$-dependent magnetic potential $m_F (V_0 + V_{\rm trap}(\mathbf{r}) )$. This is a trapping potential for the state $F=2,m_F=2$ we work with. $V_0$ corresponds to the offset magnetic field $B_0$ at the center of the magnetic trap $\mathbf{r} = (0,0,0)$. When a rf field $B_1 \cos (\omega_{\rm rf} t)$ of Rabi frequency $\Omega = g \mu_B B_1 / 2 \hbar$ --- were $g=1/2$ and $\mu_B$ is the Bohr magneton --- is applied to the atoms in the strong coupling regime, the atomic levels present an avoided crossing at the points where the rf frequency $\omega_{\rm rf}$ is resonant with the frequency difference between the $m_F$ states, that is when $V_0 + V_{\rm trap}(\mathbf{r}) = \hbar \omega_{\rm rf}$. The new potentials experienced by the dressed atoms read $m_F V(\mathbf{r})$, where:
\begin{equation}
V(\mathbf{r}) = \sqrt{(V_{\rm trap}(\mathbf{r}) - \hbar \Delta )^2 + (\hbar \Omega)^2}
\end{equation}
In this last equation, we introduced the rf detuning $\hbar \Delta = \hbar \omega_{\rm rf} - V_0$ with respect to the rf transition in the center of the magnetic trap. In this semi-classical approach, the eigenstates $\phi_{m_F}(\mathbf{r})$ corresponding to these energies are still spin states, along a direction tilted from the local direction of the magnetic field at point $\mathbf{r}$ by an angle $\theta$ where $\cos \theta = -\Delta/\sqrt{\Delta^2+\Omega^2}$ and $\sin \theta = \Omega/\sqrt{\Delta^2+\Omega^2}$. In the lab frame, the tilted eigenstate is rotating with pulsation $\omega_{\rm rf}$ around the local direction of the magnetic field. The state $\phi_{m_F}$ is identical to the uncoupled magnetic substate $-m_F$ at infinite positive detuning $\Delta$, and to $m_F$ when $\Delta$ is getting very large and negative. Note that they are eigenstates only for the internal degrees of freedom, $\mathbf{r}$ being fixed. A motional induced coupling may lead to Landau--Zener transitions between these states when $\mathbf{r}$ changes, which may be avoided by using a large enough rf coupling $\Omega$.

\begin{figure}[t]
\includegraphics[width=75mm]{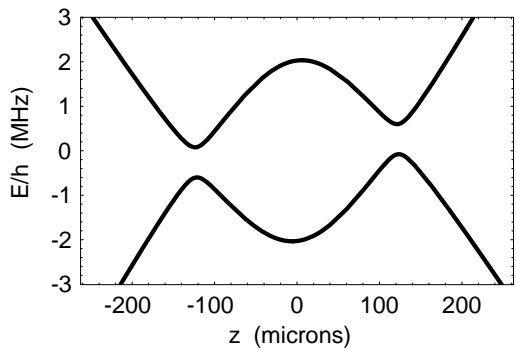}
\caption{Cut along the $z$ axis of the total potential $mgz + m_F V(\mathbf{r})$ with $m_F=2$ (upper curve) and $m_F=-2$ (lower curve). The lower state potential (corresponding to state $\phi_{-2}$) is used during evaporative cooling, where the depth is limited by the rf coupling. For the upper state ($\phi_2$), the potential minimum sits at the bottom of a shell (at $z=z_0$). Parameters are: $\Delta/2\pi=1$~MHz and $\Omega/2\pi=170$~kHz. The initial magnetic potential is the one of our QUIC trap (see text).}
\label{potcut}
\end{figure}

For atoms in the $\phi_2$ dressed state, this potential presents a minimum at the points where $V_{\rm trap}(\mathbf{r}) = \hbar \Delta$. The locus of these points is the surface where the norm of the magnetic field has a given value -- or equivalently an iso-B surface. If for instance the magnetic trapping potential is harmonic, the iso-B surfaces are ellipsoids. The atoms are thus forced to move onto an `egg shell', their motion being very limited in the direction orthogonal to the shell. This `atomic bubble' is not easily observable however: The atoms fall at the bottom of the shell due to gravity. As the typical radius of the shell increases, the atomic cloud becomes essentially two-dimensional. A cut of the potential including gravity is represented on figure~\ref{potcut}. It shows that along the vertical $z$ axis, only one side of the shell will be occupied by the atomic cloud, as soon as the temperature is lower than the energy difference between the top and the bottom of the shell.

We now will discuss the trap characteristics in the case of our experiment. The atomic cloud we start with is confined in a cigar shape QUIC magnetic trap~\cite{Esslinger98}. The long axis of the cigar is horizontal and is denoted as $x$. This is also the direction of the offset magnetic field of 1.8~G, corresponding to $V_0/h=1.3$~MHz. $y$ is the other horizontal axis along which the magnetic rf field is aligned~\cite{note1}. Finally, $z$ is the vertical axis. The oscillation frequencies at the bottom of the QUIC trap are $\omega_x/2\pi = 21$~Hz and $\omega_y/2\pi = \omega_z/2\pi = 200$~Hz. The QUIC trap is symmetric with respect to the $y=0$ plane. Initially, the atoms are confined in the upper magnetic state $F=2, m_F=2$. To load the atoms from this 3D trap into the bottom of a shell, we simply chirp a rf field from below the resonance at the center of the QUIC trap $V_0/\hbar$ to the desired value of $\Delta$. The atoms thus always stay in the upper state we start in, as soon as the Rabi frequency is large enough for them to follow adiabatically the dressed state $\phi_2$.

The cloud is located on an iso-B surface~\cite{QUICisoB}, centered at position $x_0, y_0, z_0$ where $y_0=0$ and $z_0$ is the lowest height which fulfills $V_{\rm trap}(x_0,0,z_0) = \hbar \Delta$. As $V_{\rm trap}$ depends only slightly on $x$ near the cloud position, $z_0$ is essentially given by $V_{\rm trap}(0,0,z_0) = \hbar \Delta$. The oscillation frequency in the strongly confined $z$ direction may be inferred from the Rabi frequency $\Omega$ and the vertical gradient $\alpha(z_0)$ at position $z_0$ of the trap potential, defined by: $V_{\rm trap}(0,0,z)\simeq \hbar \Delta + (z-z_0)\,\alpha(z_0)$. With these notations, the oscillation frequency in the tightly confining direction $z$ reads:
\begin{equation}
\omega_{\rm trans} = |\alpha(z_0)|\sqrt{\frac{2}{m \hbar \Omega}}
\end{equation}
In a QUIC trap, the radial gradient $\alpha$ is constant except in a small region around the minimum at $z=0$. The transverse oscillation frequency can still be monitored using the Rabi frequency $\Omega$. However, the Rabi frequency cannot be chosen arbitrarily small because of Landau--Zener losses.

The horizontal frequencies $\omega_1$ and $\omega_2$ corresponding respectively to the $x$ and $y$ directions directly depend on the local shape of the iso-B surface. In the $yz$ plane, due to the axial symmetry of the QUIC trap~\cite{QUICsymmetry}, the iso-B lines are circles and $\omega_2$ is merely the pendulum pulsation:
\begin{equation}
\omega_2 = \sqrt{\frac{g}{|z_0|}}
\end{equation}
In the $xz$ plane, the iso-B lines are very elongated due to the cigar shape of the QUIC trap and they are more complicated~\cite{QUICisoB}. $\omega_1$ is much smaller than $\omega_2$ and cannot be expressed analytically. Its order of magnitude is given by approximating the iso-B lines by ellipses:
\begin{equation}
\omega_1 \simeq \sqrt{\frac{g}{|z_0|}}\frac{\omega_x}{\omega_z}
\end{equation}

As $|z_0|$ increases with $\Delta$, $\omega_1$ and $\omega_2$ may be controlled via $\Delta$ and $\omega_{\rm trans}$ via $\Omega$; one can easily find a set of parameters such that $\omega_{\rm trans} \gg \omega_1,\omega_2$.

The experimental setup has been described elsewhere~\cite{Colombe03}. In brief, if consists of a double MOT system, the atoms being continuously loaded from an upper vapor MOT into a lower MOT sitting in a low pressure glass cell. After a transfer in the magnetic QUIC trap, the atoms are cooled down to the condensation threshold in 30~s by evaporative cooling (in the $\phi_{-2}$ dressed state). We then switch off the evaporation rf and switch on the trapping rf at a frequency below the resonance frequency in the center of the trap (1.3~MHz). This ensures that the atoms will be transferred in the $\phi_2$ state, which is connected with the bare $m_F=2$ state at very large negative detuning. $\omega_{\rm rf}$ is chirped from $2 \pi \times 1$~MHz to the desired value of the detuning $\Delta$ in typically 150~ms. The rf field is produced by a  circular antenna of axis $y$, fed by a rf synthesizer followed by either a 5~W or a 25~W amplifier. The Rabi frequency $\Omega/2\pi$ may be adjusted via a rf attenuator between 0 and 180~kHz and has been calibrated experimentally. We checked that the initial negative detuning of 300~kHz was large enough to populate only the $\phi_2$ state when switching on the rf in a 2~ms linear ramp, at $\Omega/2\pi = 180$~kHz.

\begin{figure}[t]
\includegraphics[width=70mm]{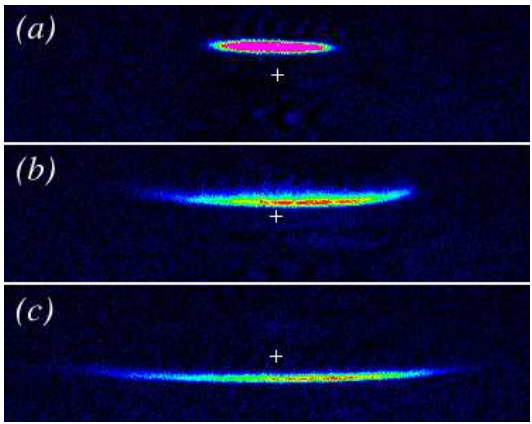}
\caption{Successive absorption images of thermal clouds in the $xz$ plane as the detuning $\Delta$ is increased. $(a)$: QUIC trap. $(b)$: $\Delta/2\pi=1.7$~MHz, $z_0=-130~\mu$m. $(c)$: $\Delta/2\pi=6.7$~MHz, $z_0=-450~\mu$m. The white cross is a fixed reference in each picture. The pictures are taken as the rf and the magnetic trap were still on. The Rabi frequency was $\Omega/2\pi= 180$~kHz. The atomic clouds contain typically $10^6$ atoms at a temperature of $1~\mu$K.}
\label{ZGtrap}
\end{figure}

The series of absorption images shown on figure~\ref{ZGtrap} illustrate the deformation of the atomic cloud as it is transferred into the egg shell trap. Starting from the QUIC trap, the cloud is translated along $z$ when $\Delta$ increases (as $z_0$ depends on $\Delta$) and is deformed at the same time: It is compressed in the $z$ direction whereas it is relaxed along $x$. The curvature of the trap, due to the iso-B surface curvature, is clearly visible. The vertical position of the cloud $z_0$ is a linear function of $\Delta$ as soon as $\Delta/2\pi$ exceeds 1~MHz, as one would expect for a constant magnetic field gradient. The measurements give access to the gradient $\alpha$ which corresponds to 225~G/cm.

In order to better characterize the egg shell trap, the oscillation frequency along $z$ was measured at a given value of $\Omega/2\pi=180$~kHz and $\Delta/2\pi=2$~MHz. The vertical oscillation frequency may be estimated to 550~Hz, knowing $\alpha$ and $\Omega$. To measure it, we excited the dipolar resonance at $\omega_{\rm trans}$ by modulating with a 5~kHz depth the detuning $\Delta/2\pi$ and thus the trap position $z_0$ for 150~ms. The vertical size of the cloud after a 10~ms time of flight was measured and plotted as a function of the excitation frequency. We found a peak centered at 600~Hz with a HWHM width of 80~Hz, in good agreement with the expected value. The small discrepancy is likely to be due to the approximate knowledge of the Rabi frequency $\Omega$.

The vertical oscillation frequency (600~Hz) corresponds to a temperature of 30~nK. As the 3D transition temperature for $10^5$~atoms in this trap is about 50~nK, pure thermal clouds confined in the adiabatic trap always remain three-dimensional. However the situation is different for a BEC where we shall compare the chemical potential $\mu$ to the vertical oscillation frequency. The atom number, the cloud position $z_0$ and the oscillation frequency $\omega_{\rm trans}$ were measured in order to estimate $\mu$. $\omega_1$ and $\omega_2$ were given a value deduced from the measured $z_0$. For an extreme detuning $\Delta/2\pi=8.7$~MHz, we have $z_0=-560~\mu$m and oscillation frequencies 600~Hz$\times$21~Hz$\times$2~Hz. For $10^5$~atoms, assuming Thomas--Fermi theory in 3D, this yields a chemical potential $\mu/h=400$~Hz which is smaller than the vertical oscillation frequency. Therefore a condensate confined in this trap would present a 2D character.

However, we did not obtain a 2D condensate in this anisotropic trap. When starting with a BEC, we observed a heating during the loading phase, which destroyed the BEC. We investigated the heating rate and lifetime of the egg shell trap using thermal gases initially prepared at $0.75~\mu$K typically in the QUIC trap. The atoms where transferred in the egg shell trap within 150~ms. After a variable holding time at a constant rf frequency (plateau) of 3~MHz in the anisotropic trap, a time-of-flight image was taken 10~ms after releasing the atoms from the trap. We measured the cloud size along $x$ and $z$. The experiment was repeated using two different schemes for producing the rf field: (i) Either we produced the rf ramp and the rf plateau with the same synthesizer (Agilent 33250A) used in a voltage controlled frequency mode (FM mode), (ii) or we used a first DDS-synthesizer (Direct Digital Synthesis, Stanford DS345) for ramping the frequency and switched to another synthesizer (Rohde \& Schwarz SML01) used at constant frequency (3~MHz) for the plateau. Note that the relative phase is not controlled at the switching.

\begin{figure}[t]
\includegraphics[width=85mm]{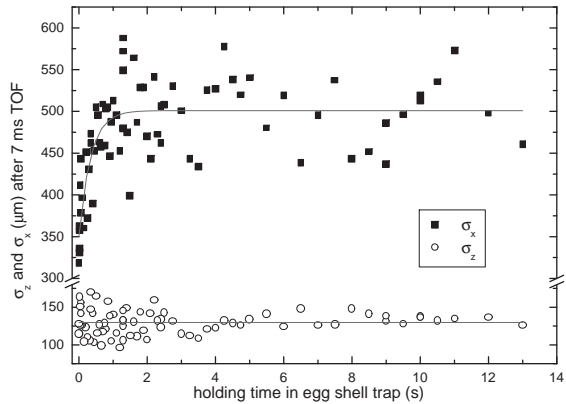}
\caption{Evolution of the cloud sizes along $x$ (filled squares) and $z$ (open circles) with the holding time in the egg shell trap after 7~ms time of flight. At this value, the expanding cloud is still anisotropic. The $x$ data are fitted with an exponential function. The deduced thermalization time is 340~ms. The equilibrium temperature measured along $z$ is $3.6~\mu$K, obtained by a fit to the $z$ data by a constant function.}
\label{heating}
\end{figure}

For both schemes, we observed a strong heating along the vertical direction during the loading rf ramp. The vertical temperature deduced from a time-of-flight measurement increased to about $4~\mu$K, whereas the horizontal temperature remained almost unchanged. We attribute this heating either to the frequency noise on the rf signal when a synthesizer is used in FM mode for scheme (i) (see below) or to the discrete frequency steps in the case of the DDS ramp (ii). During the following holding time at constant rf frequency, the heating and lifetime were very different for the two schemes. In scheme (i), the heating was about $5~\mu$K/s and the $1/e$-lifetime was limited to 360~ms. These features are clearly limited by the noise on the rf signal in FM mode: For our Agilent 33250A synthesizer used in this mode, we indeed measured a jitter at a given rf frequency during 1~s ranging from about 8~Hz FWHM for a 2~MHz modulation depth (which we use to produce a 2~MHz sweep between 1 and 3~MHz) to about 100~Hz FWHM for a 10~MHz modulation depth. These figures have to be compared to its sub-mHz spectral width if this same synthesizer were used in a fixed frequency mode. Note that these values strongly depend on the synthesizer. On the contrary, for scheme (ii) we observed a thermalization between the $x$ and $z$ directions in about 340~ms (figure~\ref{heating}) but no heating during holding time, and could obtain lifetimes as high as 4.5~s. This lifetime may be due to Landau--Zener transitions to untrapped states. In addition, the random phase jump occurring at the switching between ramping and holding rf synthesizers had a dramatic effect on the atom number; when the phase jump was close to $\pi$ (phase reversal) we observed a strong atom loss with typically five times less atoms remaining trapped as compared to quasi-continuous phase occurrences.

In conclusion, we were able to trap ultracold atoms in a new kind of anisotropic trap, produced with the same tools broadly used for obtaining a BEC, that is a magnetic trap and a rf field. This trap may be well suited for confining BECs in lower dimensions. We confined thermal clouds in this egg shell trap without noticeable heating and measured a lifetime up to 4.5~s. Defects in the rf sweep during the loading stage induced heating and prevented us from producing a 2D BEC in this trap. The heating originates from excitations along the transverse axis, either due to frequency noise of the synthetizer used in FM mode or to discrete frequency steps when using a DDS synthetizer; we currently work towards getting rid of these technical limitations and producing a BEC in the 2D regime. Another issue will then arise, as the loading phase is likely to produce excitations along the long axis $x$ of the egg shell trap, with oscillation frequency on the order of a few Hz only. To avoid these longitudinal excitations, we experimented evaporative cooling directly inside the egg shell trap using scheme (i) together with an additional rf field as proposed in \cite{Zobay01}. This could be a suitable means to achieve BEC in this trap, starting with a thermal cloud. However, this cooling mechanism could not overcome the noise-induced heating present during the holding phase in scheme (i). This evaporative method applied to scheme (ii) should give positive results.

Finally, the principle of adiabatic trapping may be associated to optical fields in order to produce new trapping geometries. In association with a red detuned vertical standing wave, the egg shell trap could be a way to load a single optical antinode \cite{Rudi}. Once the atoms fill a single antinode, a further increase of the rf detuning $\Delta$ pushes the atoms along the iso-B shell outwards the centre of the pancake. One ends up with a ring trap of about 1~mm in diameter, with high oscillation frequencies along the two transverse directions due either to the vertical standing wave or to the rf induced avoided crossing of the Zeeman sublevels. Neutral atoms were already confined in cm scale magnetic storage rings \cite{storagerings}. Our 1D ring would have no local coupling to an external atomic source, and hence no local defect. It represents a tool well adapted to the study of permanent superfluid current.

\small We gratefully acknowledge support by the R\'egion Ile-de-France (contract number E1213) and by the European Community through the Research Training Network ``FASTNet'' under contract No. HPRN-CT-2002-00304. Laboratoire de physique des lasers is UMR 7538 of CNRS and Paris 13 University.


\begin{thebibliography}{99}
\bibitem{note1} To optimize the rf coupling, the rf magnetic field has to be aligned in a direction orthogonal to the local static magnetic field, which in our experiment remains in the $xz$ plane as the rf detuning is increased.
\bibitem{QUICisoB} The iso-B lines are plotted for a QUIC trap in~\cite{Esslinger98}.
\bibitem{QUICsymmetry} The deviation from a circle in the $yz$ plane is less than 2\% for a radius of 2~mm.
\end{thebibliography}
\end{document}